\documentclass [10pt]{article}
\usepackage[T1]{fontenc}
\usepackage[final]{epsfig}
\usepackage{amssymb}
\usepackage{multicol}
\usepackage{graphics}
\usepackage{color}
\usepackage{ntimes}

\newcommand{\BE}{\begin{equation}}
\newcommand{\EE}{\end{equation}}

\begin{document}

\title{Markov chain of distances between parked cars}
\author{\textbf{Petr Seba${}^{1,2,3}$}\\
\small{${}^1$ University of Hradec Kr\'alov\'e, Hradec Kr\'alov\'e
- Czech Republic}\\
\small{${}^2$ Institute of Physics, Academy of Sciences of the
Czech Republic, Prague - Czech Republic}\\
\small{${}^3$ Doppler Institute for Mathematical
Physics and Applied Mathematics,}\\ \small{Faculty of Nuclear
Sciences and Physical Engineering,
Czech Technical University, Prague - Czech Republic}}

\normalsize

\maketitle

\begin{abstract}
We describe the distribution of distances between parked cars as a solution of
certain Markov process and
show that its solution is obtained with the help of a distributional fixed point equation.
Under certain conditions the process is solved explicitly. The resulting
probability density is compared with the actual parking data measured in the city.
\end{abstract}

We focus on the spacing distribution between cars
 parked parallel to the curb somewhere in the city center.
We will assume that the street is long enough to enable a parallel
parking of many cars. Moreover we assume that there are no driveways
or side streets in the segment of interest and that the street is
free of any kind of marked parking lots or park meters. So the
drivers are free to park the car anywhere provided
they find an empty space to do it. Finally we assume that
many cars a cruising for parking. So there are not free parking lots and a
car can park only when another car leaves the street.

The standard way to describe random parking is the continuous
version of the random sequential adsorption model known also as the
"random car parking problem" - see \cite{evans}, \cite{ca} for
review. In this model it is assumed that all cars have  the same length
$l_0$ and park on randomly chosen places. Once parked the cars do not leave the street.
This model leads to predictions that can be easily verified.
First of all it gives a relation between the mean bumper - to - bumper distance $\overline{D}$
and the car length: $\overline{D}\sim 0.337 \ l_0$. Further the probability density
 $Q(D)$ of the car distances $D$
behaves like \cite{ar},\cite{or}, \cite{yang}, \cite{mac}
\begin{equation}
Q(D)\approx -ln(D)
\label{log}
\end{equation}

To test this results real parking data were collected recently  in the center of London
\cite{rawal}. The average distance between the parked cars was  $152$  cm which fit perfectly with
the relation $\overline{D}\sim 0.337 \ l_0$ for $l_0=450$ cm.
The gap density (\ref{log}) was however fully incompatible with the observed facts.
The relation (\ref{log}) gives  $Q(D)\to \infty$ for $D\to 0$ . The
data from London showed however that in reality $Q(D)\to 0$
as $D\to 0$. Rawal and Rogers \cite{rawal} used therefore an amended version of the  model
 with a car re-positioning process (describing the car manoeuvering) to fit the data.
Later Abul-Magd \cite{abul} pointed out that the car
spacings observed in London can be well described by the Gaussian Unitary Ensemble of random
matrices  \cite{dyson},\cite{Mehta} if the parked cars are regarded as particles of a one dimensional
interacting gas. In both cases the authors assumed that $Q(D) \sim D^2$ for small $D$.
However the origin of the particular car re-positioning or
 car interaction remained unclear.

Our aim here is to show that the car parking can be understood as a
simple Markov process. The main difference to the previous approach is
that we enable the parked cars to leave the street vacating the space for a new car to parks there.
The distance distribution is obtained as a steady solution of the repeated car parking and car leaving process.

To derive the equations
we describe a car parking on a roundabout junction (we know that it is not allowed to park there,
but it simplifies the consideration).
The idea is simple: Assume that all cars have the same length $l_0$ and park on a roundabout junction with a
circumference $L$. The length L is chosen in interval $3 l_0 < L  < 4 l_0)$ so that maximally 3 cars can park there.
These cars define three spacings $D_1,D_2,D_3$ with
\begin{equation}\label{total}
  D_1+D_2+D_3+3l_0=L
\end{equation}
The parking process goes as follows: one randomly chosen car   leaves the street and
the two adjoining  lots  merge into a single one. In the second step a new car parks into the empty space and splits it
into two smaller lots.
Such fragmentation and coagulation processes were discussed intensively since they
apply for instance to the computer memory allocation  - see \cite{bertoin} for review.

Let the car leaves and the neighboring spacings - say the spacings $D_1,D_2$ - merge into a single
lot $D$
\begin{equation}\label{cdistances}
    D=D_1+D_2+l_0.
\end{equation}
When a new car parks to $D$ it splits it into
 $\tilde D_1, \tilde D_2$:
\begin{eqnarray}
 \nonumber
  \tilde D_1 &=& a(D-l_0) \\
  \tilde D_2 &=& (1-a)(D-l_0).
  \label{cnewdistances}
\end{eqnarray}
where $a\in (0,1)$ is a random variable with a probability density $P(a)$. It describes the parking preference of the driver.
We assume that all drivers have identical preferences, i.e. identical $P(a)$.
(The meaning of the variable $a$ is straightforward. For $a=0$ the car
parks immediately in front of the car delimiting the parking lot
from the left without leaving any empty space (very unworthy way to
park). For $a=1/2$ it parks directly to the center of the lot and
for $a=1$ it stops exactly behind the car on the right.)
Combining (\ref{cdistances}) and (\ref{cnewdistances}) lead to
\begin{eqnarray}
 \nonumber
  \tilde D_1 &=& a(D_1+D_2) \\
  \tilde D_2 &=& (1-a)(D_1+D_2).
  \label{cnewdistances2}
\end{eqnarray}
and the car length $l_0$ drops out. Taking the first of these equations and using the relation (\ref{total})
we finally obtain
\begin{equation}\label{total2}
     \tilde D_1 = a(L-3l_0-D_3)
\end{equation}
The pair $D_1,D_2$ is in no way particular. The same works when dealing with an arbitrary
pair of $D_k,D_l$; $k\neq l$; $k,l=1,2,3$. The last step is now easy. Hypothesize that after many steps a steady state is reached.
Then the distributions of $D_k, k=1,2,3$ are identical and the distances $D_k$ become copies of one random variable $D$.
The equation (\ref{total2}) becomes
\begin{equation}\label{markov}
D_{n+1}=a_n(L-3l_0-D_n)
\end{equation}
where $D_n$ denotes the variable $D$ at time $n$ and $a_n$ are independent identically distributed random variables
with a common probability density $P(a)$. So $D_n, n=1,2,3,...$ represents a simple Markov chain.  We are interested in the solution
of (\ref{markov})
in the limit $n\to \infty$. One can show \cite{ross}
that the limiting variable $D_\infty$ solves
the equation
$D_\infty \triangleq  a(L-3l_0-D_\infty)$. The symbol $\triangleq$
means that left and right hand sides of the equation have identical probability distributions.
The constant $L-3l_0$
can be easily scaled out. So finally the probability density $Q(D)$ of $D$ solves  the
equation
\begin{equation}\label{perpe}
D \triangleq  a-aD
\end{equation}

Distributional fixed point equations
of this type are mathematically well studied - see for instance \cite{dev},\cite{ald}, \cite{pen} although not much
is known about their exact solutions.
They are used for instance in the insurance and financial mathematics
and represent the value of a commitment to make regular payments
\cite{Goldie}. (In the actuarial mathematics these equations are called
perpetuities.)
But they arise also in the relation with recursive
algorithms such as the selection algorithm Quickselect, see e.g.
\cite{Hwang}, \cite{Mahmoud} and in many other areas.

For us the key point is  the choice of the probability distribution $P(a)$ characterizing the
parking behavior of the driver. It is reasonable to take a distribution
with $P(0)=P(1)=0$. A natural candidate is the $\beta$ distribution:
\begin{equation}\label{beta}
    P(a)=\beta(\alpha_1,\alpha_2,a)=\frac{\Gamma(\alpha_1+\alpha_2)}{\Gamma(\alpha_1)\Gamma(\alpha_2)}a^{\alpha_1-1}(1-a)^{\alpha_2-1}
\end{equation}
with $\alpha_1,\alpha_2>1$. The point is that independent $\beta$ distributed random variables fulfill a couple relations that will be helpful for us.
\cite{duf1},\cite{duf2}
\\

\textbf{Statement:}: Let $x_1, x_2$ be independent random variables. Further let  $x_1$ be $\beta(\alpha_1,\alpha_1+\alpha_2,x)$
and $x_2$ be $\beta(\alpha_3,\alpha_2,x)$ distributed. Then the variable $x_3:=x_2-x_2x_1$ has a $\beta(\alpha_3,\alpha_1+\alpha_2,x)$
distribution for all $\alpha_1,\alpha_2,\alpha_3 > 0$.
\\

Taking  $c=a$ we see that the variables $x_1$ and $x_3$ are equally distributed. In other words the above statement
leads to a solution of the  equation (\ref{perpe}) as follows:
\\

\textbf{Statement:}
\textit{Let $P(a)$ be given by (\ref{beta}). Then the probability density $Q(D)$ solving the
equation (\ref{perpe}) is given by $Q(D)=\beta(\alpha_1,\alpha_1+\alpha_2,D)$.}
\\

For parking it is natural to assume a symmetric $P(a)$: $P(a)=P(1-a)$, i.e. the drivers are not biased to park
more closely to the car adjacent from behind or from the front. The $\beta$ distribution is symmetric if $\alpha_1=\alpha_2$.
So we will take $P(a)=\beta(\alpha,\alpha,a)$ and this leads to the solution of (\ref{perpe}): $Q(D)=\beta(\alpha,2\alpha,D)$. The parameter
$\alpha$ is free and can be fitted to describe the parking data. However the observations in London \cite{rawal},\cite{abul}.
show that $Q(D)\approx D^2$ for small $D$. Such behavior means that
$\alpha=3$ since $Q(D)=\beta(\alpha,2\alpha,D)\sim D^{\alpha-1}$ for small $D$.

The variable $D$ described by the density $Q(D)$ has a mean $\overline{D}=1/3$.
To compare the distribution with the actual data we scale both of them to a mean value equal to 1. So
finally we get from the Markov process (\ref{markov}) that $Q(D)=(1/3)\beta(3,6,D/3)$. Please note that
the distribution is now fixed and does not contain any  free parameter.

\begin{figure}
\begin{center}
  \includegraphics[height=9cm,width=15cm]{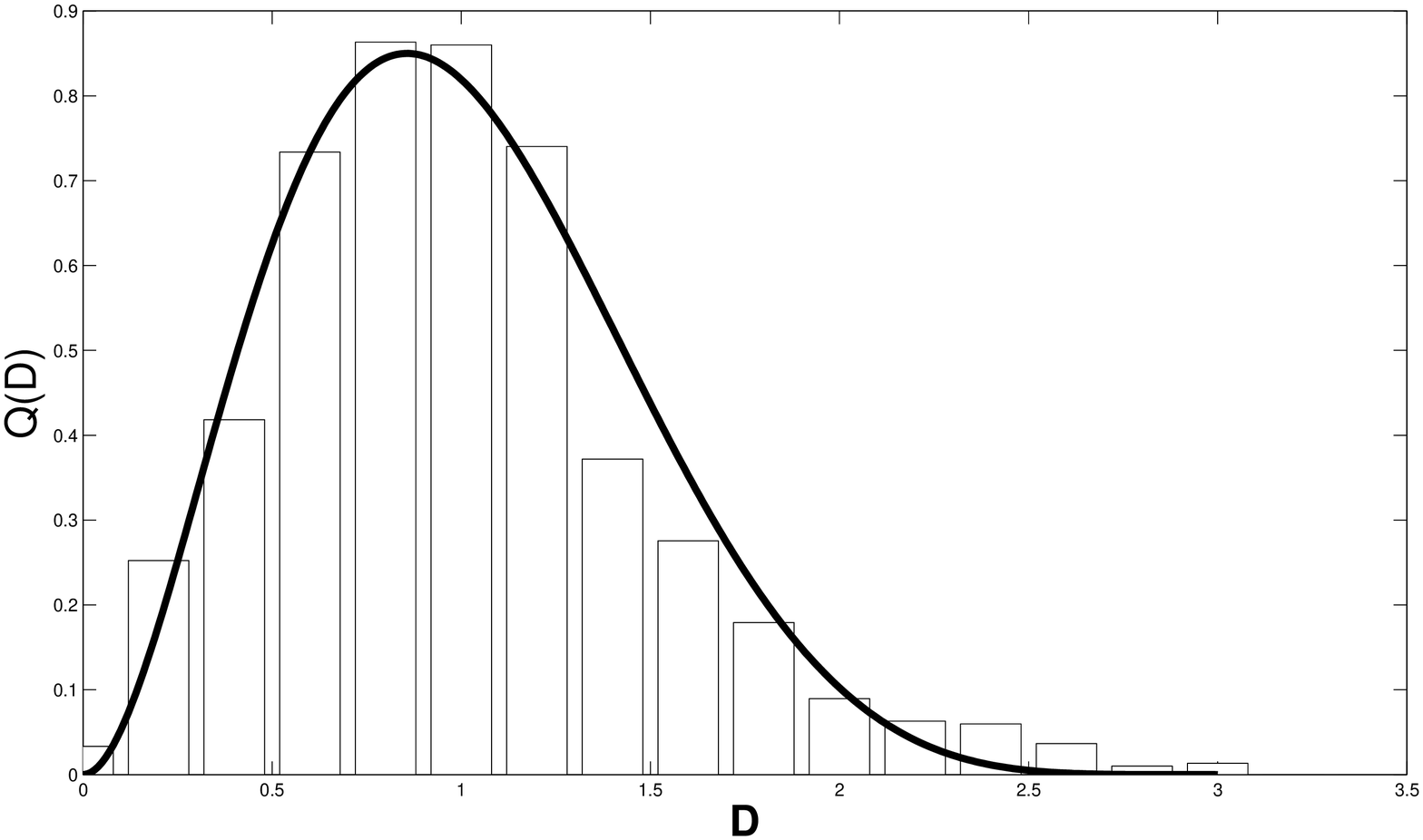}\\
\end{center}
  \caption{
The measured distance distribution for cars parking in the centre of Hradec Kralove is compared with the
solution of the equation (\ref{perpe}) with a distribution $P(a)=\beta(3,3,a)$. The measured data were scaled to a mean distance equal to 1.
  }
  \label{parking_total}
\end{figure}

To compare the prediction with the actual parking data we measured the bumper to bumper distances between the  cars parked
on two different streets in the center of Hradec Kralove (Czech Republic). Both streets were long enough and usually without any free parking place.
So a new car was able to
park there only if another car left. In addition the mean parking time was quite short so the parked cars exchanged frequently during the day.
We measured on the whole 1200 car spacings. The measured mean car distance was 140 cm.
It is slightly less then the value 154 cm reported in \cite{rawal}, but we have smaller cars in the Czech Republic.

The result is plotted on the figure \ref{parking_total} and the agreement with the prediction is reasonable. It has to be notified that the
distribution $Q(D)=(1/3)\beta(3,6,D/3)$ is quite close to the result obtained by the Gaussian Unitary Ensemble and used in \cite{abul}.
The obtained solution $Q(D)=0$ for $D>3$. This means that in the model there are not parking lots larger then 3. For the actual data (due to the
the scaling to $\overline{D}=1$) this means that the are not lots longer the 420 cm.
This is really
a favorable property, since such large lots are immediately occupied by a new car and do not persist. Such feature is however not present in the
random matrix description \cite{abul} since these distribution have long tails.

The presented model is a simplification. In reality the car do not parking on a roundabout junctions and their lengths are not equal.
But nevertheless  the simplified and solvable model leads to reasonable good agreement with the data.
To summarize: we have shown that car parking can be reasonably well described by a simple Markov process.
The limiting distribution of this process is solvable
and leads to a prediction of the car gap distribution which is in agreement
with actual data measured on the streets of the city  Hradec Kralove.

{\bf Acknowledgement:}
The research was supported by the Czech Academy of
Sciences and the Ministry of Education, Youth and Sports within the
projects A100480501 and LC06002. The help of the PhD. students Michal Musilek and  Jan Fator who collected the
car distances is gratefully acknowledged.

\end{document}